\documentclass[aps, prl, twocolumn, superscriptaddress]{revtex4}
\usepackage{graphicx}
\usepackage{amsmath}
\usepackage{amssymb}

\usepackage[colorlinks, linkcolor=blue, urlcolor=magenta, citecolor=blue]{hyperref}

\begin{document}

\title{Feedback-Induced Nonlinear Spin Dynamics in an Inhomogeneous Magnetic Field}

\author{Tishuo Wang}
\affiliation{Guangdong Provincial Key Laboratory of Quantum Metrology and Sensing, and School of Physics and Astronomy, Sun Yat-Sen University (Zhuhai Campus), Zhuhai 519082, China}
\affiliation{State Key Laboratory of Optoelectronic Materials and Technologies, Sun Yat-Sen University (Guangzhou Campus), Guangzhou 510275, China}

\author{Zhihuang Luo}
\email[]{luozhih5@mail.sysu.edu.cn}
\affiliation{Guangdong Provincial Key Laboratory of Quantum Metrology and Sensing, and School of Physics and Astronomy, Sun Yat-Sen University (Zhuhai Campus), Zhuhai 519082, China}
\affiliation{State Key Laboratory of Optoelectronic Materials and Technologies, Sun Yat-Sen University (Guangzhou Campus), Guangzhou 510275, China}

\author{Shizhong Zhang}
\affiliation{Department of Physics and Hong Kong Institute of Quantum Science and Technology, The University of Hong Kong, Hong Kong, China}

\author{Zhenhua Yu}
\email[]{huazhenyu2000@gmail.com}
\affiliation{Guangdong Provincial Key Laboratory of Quantum Metrology and Sensing, and School of Physics and Astronomy, Sun Yat-Sen University (Zhuhai Campus), Zhuhai 519082, China}
\affiliation{State Key Laboratory of Optoelectronic Materials and Technologies, Sun Yat-Sen University (Guangzhou Campus), Guangzhou 510275, China}

\begin{abstract}  
 Nonlinear effects are the root of interesting phenomena such as masers and lasers, and play a significant role in science and engineering. In spin systems, nonlinear spin dynamics is crucial for the prediction of complex dynamical behavior such as self-organizing oscillation, with applications ranging from spin masers and time crystals to precision measurement. However, when a spin system operates in a static magnetic field, how the inhomogeneity of the field affects its dynamics is a primary concern.
 Here we study the dynamics of a collection of spins with multiple Larmor frequencies for modeling a static inhomogeneous magnetic field, and reveal that due to the nonlinearity induced by a feedback scheme, the spin system exhibits much richer stable dynamical phases, including quasi-periodic orbits and chaos besides the usual limit cycles emerged in previous works. These phases are generally applicable to coupled nonlinear spin systems, even with more than two intrinsic Larmor frequencies or in continuum cases. Furthermore, we discuss their robustness against the experimental noises and the feasibility of realization in several spin systems. Our findings contribute to future observation of nonlinear dynamical phases and prospective applications in precision measurement.
\end{abstract}

\maketitle

\emph{Introduction.}---Nonlinear systems that exhibit self-organized oscillations or chaotic dynamical behavior are of great interest to physicists~\cite{Strogatz2018Nonlinear, Lin2000Resurrection, Abergel2002Chaotic}. For example, lasers (masers) are nonlinear systems that, when external pumping strength exceeds a certain threshold, begin to oscillate in phase and emit coherent lights (microwaves)~\cite{Maiman1960Stimulated,  Gordon1954Molecular, Gordon1955The, Goldenberg1960Atomic, Oxborrow2012Room, Kraus2014Room, Breeze2018Continuous, Suefke2017Para}, providing frequency references crucial for precision measurement.
To extend to low-frequency domains and operate continuously under ambient conditions without cryogenic refrigeration and high-vacuum cavities, spin maser technologies were developed to replace cavities with feedback coils~\cite{Bloom1962Principles, Chalupczak2015Alkali, Bevington2021Object}; 
the nonlinearity brought about by the feedback magnetic field generated by the coils can sustain the precession of atomic spins around an external bias magnetic field at the Larmor frequency, which corresponds to a dynamical phase of limit cycles.
The spin maser properties of noble gases have been extensively studied~\cite{Robinson1964He3, Yoshimi2002Nuclear, Jiang2021Floquet, Su2022Review}, due to their long nuclear spin coherent times protected by the closed-shell electronic configurations and high degrees of population inversion achieved via the spin-exchange collisions with auxiliary alkali-metal atoms which can be optically pumped~\cite{Walker1997Spin}. 

Continuous oscillations in spin masers can persist far beyond the transverse relaxation time $T_2$, permitting long-time coherent measurement of frequencies with greater precision~\cite{Yoshimi2002Nuclear, Inoue2016Frequency}. 
Employment of multiple species of atoms can further mitigate uncertainty arising from long-term drifts of bias magnetic field~\cite{Chupp1994Spin, Stoner1996Demonstration, Bear1998Improved, Sato2018Development, Bevington2020Dual}. These features are extremely advantageous for the precision measurement of frequency shifts, and prompt applications of spin maser technologies in the searches for permanent electric dipole moments (EDMs)~\cite{Rosenberry2001Atomic, Inoue2016Frequency}, tests of Lorentz and CPT violation~\cite{Bear2000Limit, Safronova2018Search}, and probes of dark matter~\cite{Afach2021Search, Terrano2022Comgnetometer}.   
However, in situations when spin masers utilize multiple species of atoms, in which multiple intrinsic Larmor frequencies exist as the gyromagnetic ratios of different species are different, previous works usually treat the oscillation of each species independently~\cite{Sato2018Development}. Such an approximation, though rendering agreements with experiments in certain limiting situations, hardly constitutes a complete picture of the nonlinear spin dynamics in the presence of multiple intrinsic Larmor frequencies~\cite{Liu2019Trispin}.

In this work, we study the nonlinear dynamics of a collection of spins subjected to a common feedback magnetic field and a static inhomogeneous magnetic field. The inhomogeneity of the field results in different Larmor frequencies acting on spins. We first consider a situation with binary Larmor frequencies, and find that the nonlinear spin dynamics is much richer than that for a single bias magnetic field. The resulting phase diagram consists of not only stable limit cycles that appear in previous works, but also quasi-periodic orbits and chaos. In the phase of limit cycles, all the spins, regardless of their intrinsic Larmor frequencies, manage to synchronize and sustain a sinusoidal oscillation at the central frequency. In the phase of quasi-periodic orbits, the Fourier transform of the transverse spin component peaks at multiple frequencies with equal spacing. Both limit cycles and quasi-periodic orbits exhibit robust self-organizing patterns in time and ultra-high resolution spectra in frequency, providing a wide range of potential applications in multimode spin masers, time crystals and quasi-crystals, and precision measurement. It is also possible to realize the tuning between a single-mode and multimode spin maser. In the phase of chaos, the dynamics is found to have a close resemblance to the butterfly pattern of the well-known Lorenz equations. We then generalize to more than two intrinsic Larmor frequencies and even the continuum case. We argue that limit cycles, quasi-periodic orbits, and chaos are the general phases of the nonlinear spin dynamics in the presence of multiple intrinsic Larmor frequencies. Finally, we discuss the possibility of experimental realization in spin systems such as alkali-metal atoms, noble gases, and liquid nuclear magnetic resonance.

\emph {Formalism.}---We consider a collection of $N$ spins in an inhomogeneous bias magnetic field along the $\hat{z}$-direction and a feedback field along $\hat{x}$- and $\hat{y}$-directions, i.e., $\mathbf{B} = (B_x^{\rm fb}, B_y^{\rm fb}, B_0^j)$, where $B_0^j$ is assumed to be a local bias field experienced by the $j$-th spin for simulating the inhomogeneous magnetic field. Therefore, the dynamics of spin polarization vector $\mathbf P_j=(P_{j,x},P_{j,y},P_{j,z})$ can be described by a set of nonlinear Bloch equations
\begin{align}
\frac{d P_{j,x}}{dt}=&\omega_j P_{j,y}-\gamma B_y^{\rm fb} P_{j,z}-\frac{P_{j,x}}{T_2 },\label{blochx}\\
\frac{d P_{j,y}}{dt}=&-\omega_j P_{j,x}+\gamma B_x^{\rm fb} P_{j,z}-\frac{P_{j,y}}{T_2 },\label{blochy}\\
\frac{d P_{j,z}}{dt}=&\gamma B_y^{\rm fb} P_{j,x}-\gamma B_x^{\rm fb} P_{j,y}-\frac{P_{j,z}-P_0}{T_1}.
\label{blochz}
\end{align}
Here $P_0$ denotes the equilibrium polarization, $\gamma$ is the gyromagnetic ratio of spin, $T_1$ and $T_2$ stand for the (effective) spin longitudinal and transverse relaxation times. The Larmor frequency for each spin $\omega_j = \gamma B_0^j$ can be different due to the inhomogeneity of the bias field. All spins are subjected to a common feedback field: $B_x^{\text{fb}}(t)=\frac{\alpha}{\gamma}\overline P_{y}(t)$ and $B_y^{\text{fb}}(t)=-\frac{\alpha}{\gamma}\overline P_{x}(t)$ that are proportional to the transverse components of the average spin polarization defined as $\overline{\mathbf P}(t)\equiv (1/N)\sum_{j=1}^N \mathbf P_j(t)$ ~\cite{Yoshimi2002Nuclear}. This feedback field can be realized by a circuit which measures the transverse components $\overline{P}_x(t)$ and $\overline{ P}_y(t)$ and feed the measurement results into pairs of Helmholtz coils with the amplification factor $\alpha (>0)$ \cite{Bloom1962Principles, Chalupczak2015Alkali, Bevington2021Object}. In the case when the bias magnetic field is homogeneous, i.e., $\omega_j$ are identical, it has been shown that 
the no signal fixed point $\mathbf P_j=\mathbf P_{\rm NS}\equiv(0, 0, P_0)$, corresponding to zero transverse spin polarization, becomes unstable once $\alpha$ exceeds the critical value $\alpha_c\equiv1/T_2P_0$. For $\alpha>\alpha_c$, a new stable limit cycle solution emerges for which a non-zero transverse spin polarization processes at the identical Larmor frequency, corresponding to the onset of maser \cite{Yoshimi2002Nuclear}. 

\emph{Binary Larmor frequencies.}---To investigate the effects of an inhomogeneous bias magnetic field, we start with considering the situation of binary Larmor frequencies in which half of the spins are subject to  $\omega_1=\omega_c+\epsilon/2$ and the other half $\omega_2=\omega_c-\epsilon/2$. Without loss of generality, we assume $\epsilon>0$. In this case, we only need to consider two representative spins $\mathbf P_1$ and $\mathbf P_2$, equivalent to taking $N=2$ in Eqs.~(\ref{blochx})-(\ref{blochz}). The coupled Bloch equations in this binary case are six-dimensional; trajectories of the long-time stable solutions obtained by numerically solving the equations would be rather difficult to analyze and visualize. In experiments, however, the spins are usually prepared in the same initial state, i.e., $\mathbf P_1(t=0)=\mathbf P_2(t=0)$. Starting from this initial condition, we can prove that subsequent evolution maintains $P_{1,z}=P_{2,z}$ and $|P_{1,T}|=|P_{2,T}|$ at any time $t$~\cite{SM} with $P_{j,T}\equiv P_{j,x}+iP_{j,y}$ the complex transverse component. These two features allow us to reduce the dimension of the nonlinear system. 
The second feature ensures that the mean component $\overline P_{T}\equiv (P_{1,T}+P_{2,T})/2$ is perpendicular to the difference $\Delta P_T\equiv P_{1,T}-P_{2,T}$, which prompts us to parameterize $\overline P_T=Ae^{i\theta}/2$ and $\Delta P_T=B e^{i(\theta+\pi/2)}$ with real amplitudes $A$ and $B$ and phase angle $\theta$. Consequently, the Bloch equations are reduced to the form
\begin{align}
\frac{d A}{dt}=&\alpha \overline P_z A+\epsilon B/2-A/T_2 ,\label{a}\\
\frac{d B}{dt}=&-\epsilon A/2-B/T_2 ,\label{b}\\
\frac{d \overline P_z}{dt}=&-\alpha A^2/4-(\overline P_z - P_0)/T_1,\label{pz}\\
\frac{d\theta}{dt}=&-\omega_c.\label{theta}
\end{align}
Since the phase angle can be readily solved as $\theta(t)=-\omega_c t+\phi$ with $\phi$ an arbitrary phase, the remaining dynamical system is reduced to three dimensions regarding $\{A, B, \overline P_z\}$. In the following, we will use dimension reduction to simplify our analysis. For initial conditions other than $\mathbf P_1(t=0)=\mathbf P_2(t=0)$, see discussions in supplementary materials for details~\cite{SM}.

\begin{figure}
	\centering
	\includegraphics[width = 1 \linewidth]{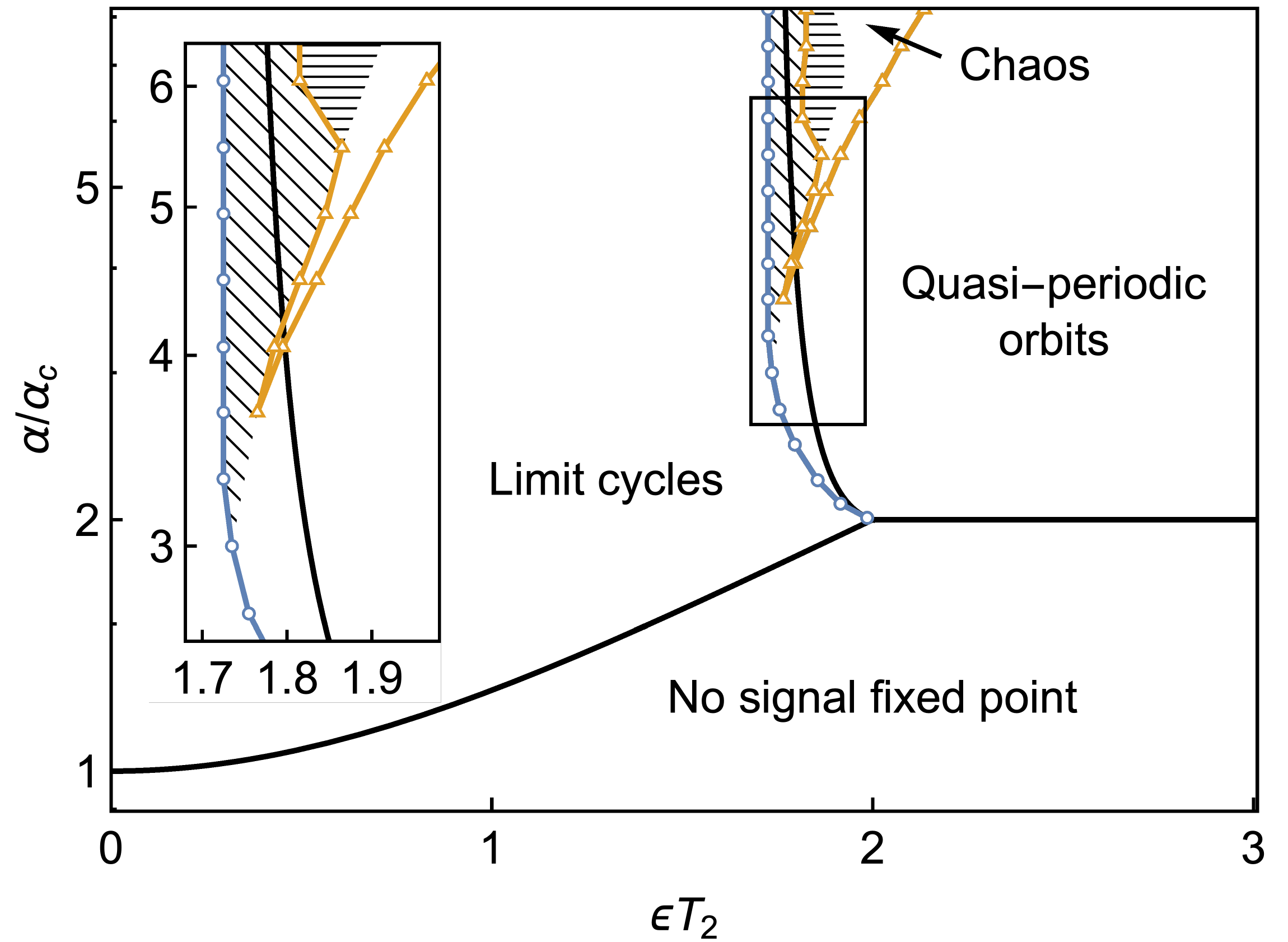}
	\caption{Stable phase diagram consisting of limit cycles, quasi-periodic orbits, and chaos, beyond the no signal fixed point. The stable region boundaries represented by the three black solid lines are (i) $\alpha/\alpha_c = 1 + (\epsilon T_2 /2)^2$ for $\epsilon T_2  < 2$, (ii) $\alpha/\alpha_c =2$ for $\epsilon T_2  \geq 2$, and (iii) $\alpha/\alpha_c = 3y/2+(1-d)/2(y-d)$ for $\min(1,d)<y<\max(1,d)$, where $y\equiv (\epsilon T_2 /2)^2$ and $d\equiv T_2/T_1$. The symbols of triangle and circle mark the stable region boundaries determined numerically. The lines linking the symbols are guide for the eye. There exist overlaps of the stable regions. The two features, $P_{1,z}=P_{2,z}$ and $|P_{1,T}|=|P_{2,T}|$, are found maintained by the stable trajectories numerically solved, except for the quasi-periodic orbits in the diagonally slashed area and for the chaos in the horizontally slashed area \cite{SM}. The inset shows a zoom-in of the boxed part. }
	\label{fig: stablediagram}
\end{figure}

\begin{figure*}
	\centering
	\includegraphics[width = 0.95\linewidth]{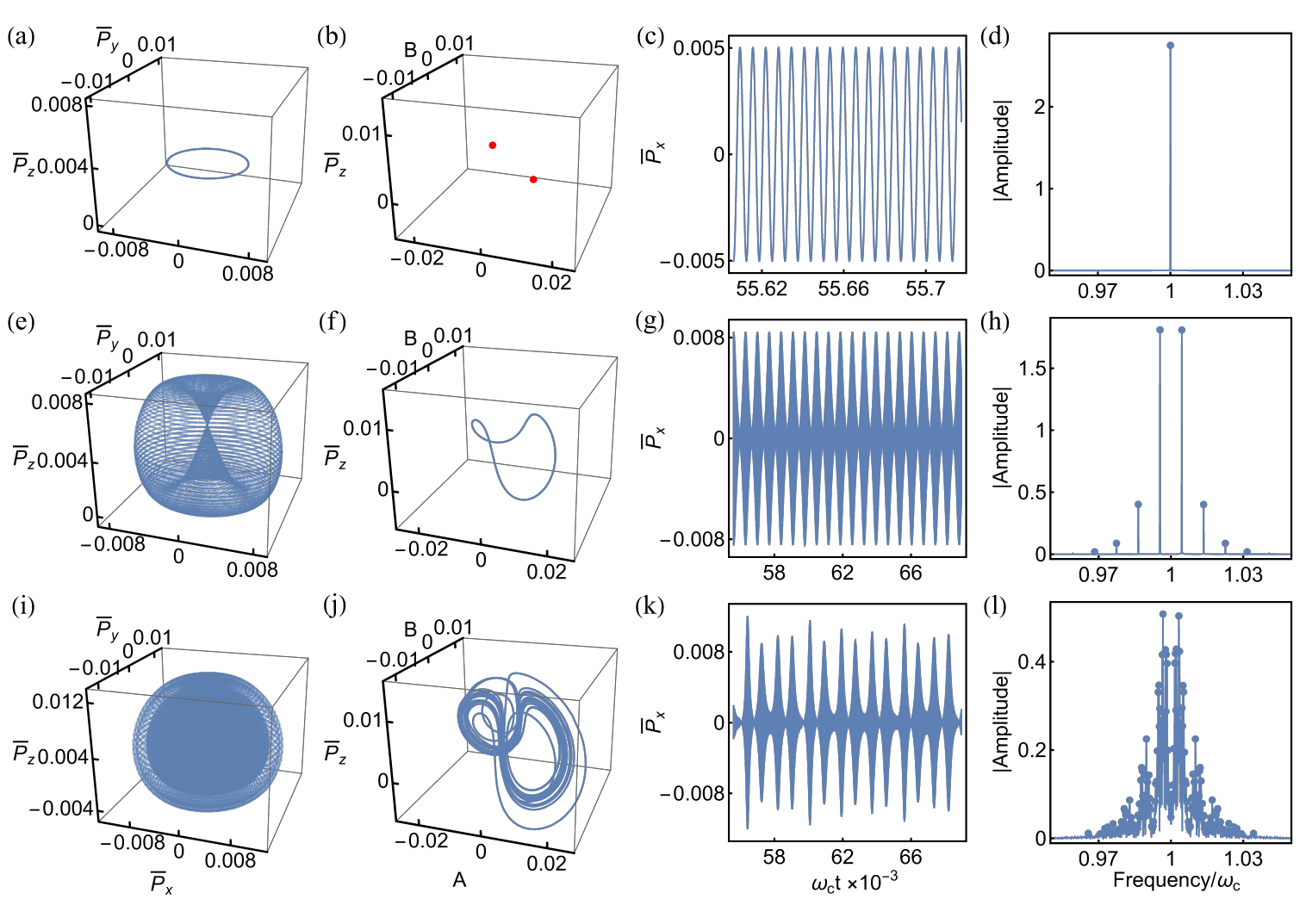}
	\caption{Stable dynamical behaviors of limit cycles (a)-(d), quasi-periodic orbits (e)-(h), and chaos (i)-(l). The first column of the graphs plots the trajectories of the overall mean polarization $\overline{\mathbf P}(t)$. The second column is the corresponding trajectories in the phase space of $\{A,B,\overline P_z\}$; the limit cycles condense into twin fixed points $(A_{\rm LC,\pm}, B_{\rm LC,\pm},\overline{P}_{z,\rm LC})$, the quasi-periodic orbits unify into a limit cycle solution, and the chaotic trajectories are reminiscent of the renowned Lorenz equations. The third column shows the time series of $\overline P_x(t)$ and the fourth is the absolute values of the Fourier transform amplitudes of $\overline P_x(t)$ in arbitrary units. Here $\alpha/\alpha_c=7.8$, and $\epsilon T_2 =1, 3, 2$ are taken for plotting the (a)-(d), (e)-(h), and (i)-(l), respectively.  }
	\label{fig: phaseportrait}
\end{figure*}

\emph{Stability diagram.}---Figure \ref{fig: stablediagram} represents the stable dynamical phases of binary Larmor frequencies.
When $\alpha$ is sufficiently large, the system is found to exhibit rich phases: limit cycles for relatively small $\epsilon$, quasi-periodic orbits for relatively large $\epsilon$, and chaos in between. The boundaries of the stable dynamical phases marked by the solid lines in Fig. \ref{fig: stablediagram} are obtained analytically via the linear stability analysis~\cite{SM}, whereas those marked by discrete symbols are determined numerically by taking the parameters $P_0=0.0174833$, $1/T_1=0.11348$ Hz, $1/T_2=0.148$ Hz, and $\omega_c/2\pi=8.85$ Hz~\cite{Chen2020Rapid}. 

(a) Limit cycles. In a dynamical system, a closed trajectory in the phase space constitutes a limit cycle solution if there exists at least another trajectory spiraling into it either as time $t\to\infty$ or as time $t\to-\infty$; the former is called stable as the latter unstable \cite{Strogatz2018Nonlinear}. The stable limit cycle solution is also called the masing phase. When $\epsilon=0$, our system reduces to the previously studied one-species spin masers~\cite{Yoshimi2002Nuclear, Inoue2016Frequency}. 
Figure~\ref{fig: stablediagram} shows that the stable limit cycle phase extends into the region where $\epsilon\neq 0$. However, to escape from the trivial phase, one needs $\alpha/\alpha_c> f(\epsilon T_2)$ with $f(x)\equiv1+(x/2)^2$; the larger $\epsilon$ is, the larger threshold value of the gain $\alpha$ is required. The limit cycle solution can be derived by setting the left-hand sides of Eqs.~(\ref{a})-(\ref{pz}) to zero. We find a pair of nontrivial twin fixed points: $A_{\rm LC,\pm}=\pm 2\{P_0[\alpha-\alpha_cf(\epsilon T_2)]/T_1\}^{1/2}/\alpha$, $B_{\rm LC,\pm}=-\epsilon T_2 A_{\rm LC,\pm}/2$ and $\overline P_{z,{\rm LC}}=f(\epsilon T_2)/\alpha T_2$, which are equivalent to the limit cycle solution of Eqs.~(\ref{blochx})-(\ref{blochz}), i.e., $P_{j,T}= [1+i (-1)^{j}\epsilon T_2/2]\overline P_T$, $P_{1,z}=P_{2,z}=\overline P_{z,{\rm LC}}$ with $\overline P_T=e^{-i(\omega_c t-\phi)}|A_{\rm LC,\pm}|/2$; the pair of the twin fixed points correspond to two periodic orbits of the limit cycles whose phase angles $\phi$ differ by $\pi$.  

The limit cycle phase is stable in the region left to the line $\alpha/\alpha_c=3y/2+(1-d)/2(y-d)$ for $\min(1,d)<y<\max(1,d)$ with $y\equiv (\epsilon T_2/2)^2$ and $d\equiv T_2/T_1$ \cite{SM}; the line starts at the point $\epsilon =2/T_2$ and $\alpha=2\alpha_c$.
 Across the line, a Hopf bifurcation occurs, rendering the limit cycles unstable~\cite{SM}. 
The stable dynamical behaviors of limit cycles are given in Figs.~\ref{fig: phaseportrait}(a)-\ref{fig: phaseportrait}(d). The sinusoidal oscillation of $\overline P_T(t)$ gives rise to a signal at the masing frequency $\omega_s=\omega_c$, and the natural linewidth ($\propto 1/T_2$) disappears in the feedback scheme. It is very important to suppress the uncertainty of the masing frequency and achieve higher precision in measurement~\cite{Terrano2022Comgnetometer}.

(b) Quasi-periodic orbits. In contrast to limit cycles, a quasi-periodic motion of a dynamical system consists of two or more incommensurable frequencies such that the corresponding trajectory is not closed \cite{Strogatz2018Nonlinear}.
In the region of $\alpha> 2\alpha_c$ for $\epsilon T_2>2$, the stable solution of our system turns out to be quasi-periodic orbits. This can be understood in the large $\epsilon$ limit. Since $\epsilon$ is much larger than any other energy scales, to lowest order, $\mathbf P_1$ and $\mathbf P_2$ precess with frequencies $\omega_1$ and $\omega_2$ respectively. 
The previous treatment approximating spin oscillations of different species independently corresponds to this limit \cite{Sato2018Development}.
Since generally the ratio $\omega_1/\omega_2$ is irrational, $\overline{\mathbf P}(t)$ shall be quasi-periodic.
Going beyond the lowest order does not change the nature of quasi-periodic orbits.

In the phase of quasi-periodic orbits, the trajectories of $\overline {\mathbf P}(t)$ look rather dense as shown in Fig.~\ref{fig: phaseportrait}(e). 
However, once represented in terms of $\{A,B,\overline P_z\}$, the quasi-periodic orbits unify into a limit cycle solution as shown in Fig.~\ref{fig: phaseportrait}(f). Note that this limit cycle solution generally has a three-dimensional configuration, and its period $\tau$, while independent of $\omega_c$, changes with $\alpha$ and $\epsilon$~\cite{SM}. 
Given Eqs.~(\ref{a}) to (\ref{pz}) are invariant under the transform $\{A,B,\overline P_z\}$ to $\{-A,-B,\overline P_z\}$, the Fourier transform amplitude of $\overline P_x (t)$ or  $\overline P_y (t)$ peaks at regular frequencies $\omega_c+2\pi (2 n+1)/\tau$ with integer $n$ as shown in Fig.~\ref{fig: phaseportrait}(h). The quasi-periodic orbit phase found in our spin system provides a realization for multi-mode excitations. Such multi-mode excitations are useful to eliminate uncorrelated noise in the frequency domain \cite{Jiang2022Floquet}. 

(c) Chaos. High sensitivity to initial conditions is a hallmark of chaotic dynamical motions \cite{Strogatz2018Nonlinear,parker2012practical,sandri1996numerical}. Interacting magnetic systems can generally exhibit chaotic motions under appropriate conditions \cite{Lin2000Resurrection, Abergel2002Chaotic}. In our system, the effective interaction is introduced by the feedback field $B_x^{\rm fb}(t)$ and $B_y^{\rm fb}(t)$. In Fig.~\ref{fig: stablediagram}, we locate the stable region of the chaos phase largely in between the limit cycle phase and the quasi-periodic orbit phase. The exact locations of the chaos phase depend sensitively on the parameters of the system \cite{Strogatz2018Nonlinear}. In contrast to Fig.~\ref{fig: phaseportrait}(i), when the chaotic trajectories is presented in terms of $\{A,B,\overline P_z\}$ as in Fig.~\ref{fig: phaseportrait}(j), a butterfly pattern, reminiscent of the well-known Lorenz equations, emerges. Figure \ref{fig: phaseportrait}(l) shows that the Fourier transform of the transverse amplitude in the chaotic regime peaks irregularly. 
An alternative quantitative way to distinguish chaos from other stable orbits, e.g, quasi-periodic orbits, is via the Lyapunov exponents, which essentially quantify the rates how fast deviations in initial conditions can grow; the largest Lyapunov exponent is positive for the former and zero for the latter. The exponent can be either numerically calculated from a known dynamical system \cite{parker2012practical,sandri1996numerical}, or estimated from a time series of data collected in experiment \cite{rosenstein1993practical,SM}.

\begin{figure}
	\centering
	\includegraphics[width=0.48\textwidth]{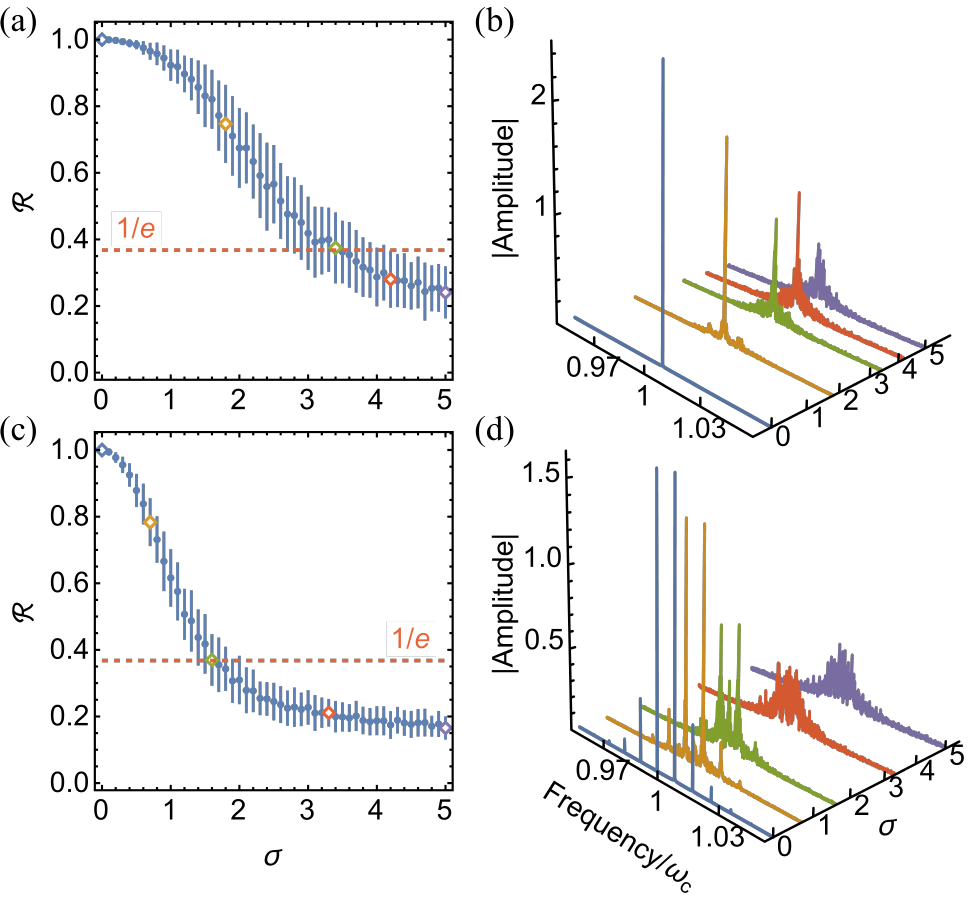}
	\caption{Robustness of the limit cycles (top row), quasi-periodic orbits (bottom row) against the fluctuations of the feedback field. The first column of the graphs plot the function $\mathcal R$ averaged over $100$ individual simulations; the error bars therein represent the corresponding standard deviations. A cut-off is set where $\mathcal R$ reduces to $1/e$.
	The second column shows the single shots of the absolute values of the Fourier transform amplitudes of $\overline P_x(t)$ for $\sigma$ representative across the cut-off. Parameters taken here are $\alpha/\alpha_c=7.8$, $\epsilon T_2=1$ for (a) and (b), and $\alpha/\alpha_c=7.8$, $\epsilon T_2=3$ for (c) and (d).}
	\label{fig: probust2spins}
\end{figure}

Due to the nonlinear nature of our dynamical system, the predicted limit cycle, quasi-periodic orbit, and chaotic phases are robust against experimental perturbations. For instance, to evaluate the effects of fluctuations from the feedback mechanism, we consider two fluctuating fields $f_\sigma(t)$ and $g_\sigma(t)$ appearing in the feedback fields as $B_x^{\text{fb}}=[\alpha\overline P_{y}+f_\sigma(t)]/\gamma$ and $B_y^{\text{fb}}=[-\alpha\overline P_{x}+g_\sigma(t)]/\gamma$.
At any time $t$, $f_\sigma(t)$ and $g_\sigma(t)$ sample randomly from the range $[-\sigma,\sigma]$. Under the influence of the fluctuations with strength $\sigma$, the time series of stable dynamics denoted by $q_\sigma(t)$ can be different from $q_0(t)$ for $\sigma=0$. To quantify the robustness of the dynamical phase corresponding to $q_0(t)$, we define the function
\begin{align}
\mathcal R & = \frac{\int  |\tilde q_0(\omega)\tilde q_\sigma(\omega)| d\omega/2\pi}{\sqrt{\int |\tilde q_0(\omega)|^2 d\omega/2\pi \int |\tilde q_\sigma(\omega')|^2 d\omega'/2\pi}}.
	\label{eq: robust}
\end{align}
where $\tilde q_\sigma(\omega)$ is the Fourier transform of $q_\sigma(t)$. 
Figure~\ref{fig: probust2spins} shows the robustness of the limit cycles and quasi-periodic orbits calculated from the time series of $\overline P_x(t)$ with and without the fluctuations of the feedback field. We can see that the cut-off noise strength $\sigma_{\rm cut}\approx 3$ and $\approx 1.5$ for the limit cycle and the quasi-periodic orbit respectively, at which $\mathcal R$ decreases to $1/e$; single-shot spectra of $\overline P_x(t)$ representative across the cut-off are shown in Fig.~\ref{fig: probust2spins}(b) and~\ref{fig:  probust2spins}(d). 
Figure \ref{fig: probust2spins} shows that the limit cycles of a single peak in the frequency domain seem more robust than the quasi-periodic orbits, which may be due to the fact that the latter are subject to the noise at their \emph{multiple} peaks simultaneously. Other practical matters relevant to the experiment such as 
glitches in setups, and fluctuations and drifts in the bias field are also considered in supplementary materials. 

\emph{Discussions.}---We have studied the nonlinear spin dynamics with binary Larmor frequencies, and analyzed in detail its rich dynamical phases induced by a feedback scheme. As the difference of binary Larmor frequencies increases, the stable solution of the limit cycles persists until it is replaced by the quasi-periodic orbits and chaotic dynamics. 
Due to the nonlinearity of the system, the self-sustained oscillations generated by the limit cycles and the quasi-periodic orbits are independent of the initial conditions and exhibit impressive robustness against experimental perturbations \cite{SM}. These nonlinear oscillations are expected to sustain for arbitrary lengths of time, which are appropriate for the long-term measurement of frequency shifts such as the search for EDMs \cite{Inoue2016Frequency, Safronova2018Search}, the advances of single-mode and multimode spin masers, and can be further considered as realizations of time crystals and quasi-crystals \cite{Phatthamon22,wu2023observation,PhysRevLett.131.143002,ding2023ergodicity,Autti2018,Giergiel2018,Huang2018Symmetry-breaking}. 



Our formalism can be further generalized to the cases with more than two intrinsic Larmor frequencies or in a continuum limit where the inhomogeneous Larmor frequencies follow a distribution $\rho(\omega)$ \cite{SM}. We find that limit cycles, quasi-periodic orbits and chaos continue to be the generic phases of the dynamics of the coupled spins, even when the feedback magnetic field is applied only in one direction \cite{Jiang2021Floquet,SM}.
Our findings can be demonstrated experimentally in spin systems such as alkali-metal atoms, noble gases, and liquid nuclear magnetic resonance. We can employ a species of atomic spins (e.g., K, Rb, Cs) or nuclear spins (e.g., $^3$He, $^{21}$Ne, $^{129}$Xe) applied in an inhomogeneous magnetic field \cite{SM} or multiple species of spins co-located in a bias field. Co-located spin masers consisting of two species of atoms, for example, ${}^{129}$Xe and ${}^{131}$Xe, are employed to mitigate the frequency instability due to the magnetic field and cell temperature drifts \cite{Sato2018Development}. Our study indicates that these co-located nuclear spin masers can operate in the quasi-periodic orbit regime. 


\emph{Acknowledgements.}---We thank Xiaodong Li and Long Wang for their discussions on numerics. TW and ZY are supported by the National Natural Science Foundation of China Grant No.12074440, and Guangdong Project (Grant No.~2017GC010613). ZL is supported by the National Natural Science Foundation of China Grant No. 
11805008, Fundamental Research Funds for the Central Universities, Sun-Yat-Sen University (Grant No. 23qnpy63), and Guangdong Provincial Key Laboratory (Grant No. 2019B121203005), and Guangdong Basic and Applied Basic Research Foundation (Grant No. 2024A1515011406). SZ is supported by grants from the Research Grants Council of the Hong Kong Special Administrative Region, China (HKU 17304719, HKU C7012-21GF, and a RGC Fellowship award HKU RFS2223-7S03).

\bibliographystyle{apsrev4-1}
\bibliography{Ref_maser}

\end{document}